\documentclass[aps,prl,superscriptaddress,reprint,floatfix,showpacs]{revtex4-1}
\usepackage{graphicx}
\usepackage{dcolumn}
\usepackage{bm}
\usepackage{natbib}
\usepackage{longtable}

\begin{document}

\title{Observation of magnetically hard grain boundaries in
double-perovskite Sr$_{2}$FeMoO$_{6}$}

\author{Y. Takahashi}
\email{takahashi@wyvern.phys.s.u-tokyo.ac.jp}
\affiliation{Department of Physics, University of Tokyo, Bunkyo-ku,
Tokyo 113-0033, Japan}
\author{V. K. Verma}
\affiliation{Department of Physics, University of Tokyo, Bunkyo-ku,
Tokyo 113-0033, Japan}
\author{G. Shibata}
\affiliation{Department of Physics, University of Tokyo, Bunkyo-ku,
Tokyo 113-0033, Japan}
\author{T. Harano}
\affiliation{Department of Physics, University of Tokyo, Bunkyo-ku,
Tokyo 113-0033, Japan}
\author{K. Ishigami}
\affiliation{Department of Complexity Science and Engineering, University of Tokyo, Kashiwa-shi, Chiba 277-8561, Japan}
\author{K. Yoshimatsu}
\affiliation{Department of Physics, University of Tokyo, Bunkyo-ku,
Tokyo 113-0033, Japan}
\author{T. Kadono}
\affiliation{Department of Physics, University of Tokyo, Bunkyo-ku,
Tokyo 113-0033, Japan}
\author{A. Fujimori}
\affiliation{Department of Physics, University of Tokyo, Bunkyo-ku,
Tokyo 113-0033, Japan}

\author{A. Tanaka}
\affiliation{Department of Quantum Matter, ADSM, Hiroshima University,
Higashi-Hiroshima 739-8530, Japan}

\author{F.-H. Chang}
\author{H.-J. Lin}
\author{D. J. Huang}
\author{C. T. Chen}
\affiliation{National Synchrotron Radiation Research Center (NSRRC),
Hsinchu 30076, Taiwan}

\author{B. Pal}
\author{D. D. Sarma}
\affiliation{Solid State and Structural Chemistry Unit, Indian Institute
of Science, Bangalore 560 012, India}
\date{\today}

\begin{abstract}
Unusual low temperature magneto-resistance (MR) of ferromagnetic Sr$_{2}$FeMoO$_{6}$ polycrystals has
been attributed to magnetically hard grain boundaries which act as spin
valves. We detected the different magnetic hysteresis curves for the
grains and the grain boundaries of polycrystalline
Sr$_{2}$FeMoO$_{6}$ by utilizing the different probing depths of the different detection modes of x-ray absorption spectroscopy (XAS) and x-ray
magnetic circular dichroism (XMCD), namely, the total electron yield
(TEY) mode (probing depth $\sim$5 nm) and the total
fluorescence yield (TFY) mode (probing depth $\sim$100 nm). At
20 K, the magnetic coercivity detected in the TEY mode
($H_{\rm c,TEY}$) was several times larger than that in the
TFY mode ($H_{\rm c,TFY}$), indicating harder ferromagnetism of the grain
boundaries than that of the grains. At room temperature, the grain
boundary magnetism became soft and $H_{\rm c,TEY}$ and $H_{\rm c,TFY}$ were nearly the
same. From line-shape analysis of the XAS and XMCD spectra, we found
that in the grain boundary region the ferromagnetic component is dominated by Fe$^{2+}$ or
well-screened signals while the non-magnetic component is dominated
by Fe$^{3+}$ or poorly-screened signals.
\end{abstract}


\maketitle

Colossal magneto-resistance (CMR) \cite{Jonker1950Ferromagnetic-c}, a giant decrease of
the electrical resistivity under a magnetic field, is a remarkable
property that can be applied to spintronics devices. CMR has been
observed in many complex oxides of Mn such as La$_{1-x}$Ca$_{x}$MnO$_{y}$
\cite{Jin1994Thousandfold-Ch}, La$_{1-x}$Ba$_{x}$MnO$_{y}$ \cite{Helmolt1993Giant-negative-}, and Tl$_{2}$Mn$_{2}$O$_{7}$ \cite{Subramanian1996Colossal-Magnet}. However, it needs high magnetic
fields in the 1 T range, making the pravtical use of these materials
difficult. On the other hand, low-field grain-boundary magneto-resistance has been discovered in the pyrochlore-type Tl$_{2}$Mn$_{2}$O$_{7}$
\cite{Hwang1997Low-field-magne}. The polycrystalline double perovskite
Sr$_{2}$FeMoO$_{6}$ (SFMO), which has a high Curie temperature $T_{{\rm
C}}=415$ K, has also been discovered to show CMR at low
magnetic field and room temperature
\cite{Kobayashi1998Room-temperatur} and therefore has become one of
the most promising materials for the application of CMR. Because the
same low-field magneto-resistance (LFMR) has not been observed in single
crystalline samples \cite{Tomioka2000Magnetic-and-el}, tunneling
magneto-resistance (TMR) between ferromagnetic grains via insulating
grain boundaries has been proposed as a mechanism of the LFMR of
SFMO \cite{Tomioka2000Magnetic-and-el}. \\
\indent However, a recent study of polycrystalline
SFMO has shown unusual magneto-resistance (MR) that cannot be explained
by the conventional TMR mechanism, according to which the peak MR should
occur at a magnetic
field which coincides with the magnetic coercivity ($H_{{\rm c}}$)
\cite{Ohno2001Spin-dependent-}. At low temperatures, the MR peak of SFMO occurs
at magnetic fields ($H_{{\rm c,MR}}$) about 6 times larger than
$H_{{\rm c}}$ \cite{Sarma2007Intergranular-M}, while such a difference between $H_{\rm c}$ and $H_{{\rm c,MR}}$
disappears at high temperatures \cite{Ray2011Origin-of-the-u}. It has been suggested
\cite{Sarma2007Intergranular-M} that the unusual MR in SFMO is controlled by the spin polarization of grain-boundary
regions acting as spin valves, and is called spin-valve-type MR
(SVMR), in contrast to the conventional MR in which the tunnel barriers are
non-magnetic. If the $H_{\rm c}$ of the magnetic tunnel barrier is
larger than that of the grains, the discrepancy between $H_{\rm c}$ and
$H_{{\rm c,MR}}$ can be understood. However, it is difficult to
differentiate between the magnetic states of
the grains and the grain boundaries using conventional magnetization
measurements.\\
\indent In this work, we have investigated the different coersivities between the grains and the grain boundaries by using x-ray magnetic
circular dichroism (XMCD) at the Fe $L_{\rm 2,3}$ edges. We have
utilized the different probing depths of different detection modes,
namely, the total electron yield (TEY,
probing depth $\sim$5 nm) and total fluorescence yield (TFY, probing
depth $\sim$100 nm) modes.\\
\indent Polycrystalline samples were prepared by arc melting and annealing in 2\% hydrogen and 98\% argon atmosphere at 1250 $^{\circ}$C
\footnote{This sample is similar to sample C in Ref \cite{Sarma2007Intergranular-M}. Details of the sample
preparation are given in Ref \cite{Sarma2007Intergranular-M}.}. Electron microscopy observation
showed that the
average grain size was 5 - 10
$\mu$m. The grain
boundaries are expected to have the thickness of order $\sim$1 nm. X-ray absorption
spectroscopy (XAS) and XMCD measurements
were carried out at the Dragon beam line 11A of National Synchrotron Radiation Research
Center (NSRRC), Taiwan. The monochromator resolution was $E/\Delta E>10000$ and
the degree of circular polarization of x-rays was $\sim$60\ \%. Angle
between the incident light and the magnetic field was 30$^\circ$, and
magnetic fields (from -1 T 
to 1 T) were applied parallel to the sample surface. The measurement
temperatures were 20 and 300 K. The samples were scraped {\it in situ} by a diamond file under an ultrahigh
vacuum of $\sim$$1\times10^{-9}$ Torr. When sintered
polycrystalline samples are scraped, if intergranular fracture
dominates intragranular one, the scraped surface would largely consist of
grain boundaries. Because the thickness of the grain
boundaries is expected to be of order a few nm, it is comparable to the probing depth of
the TEY mode. Therefore, spectra taken in the TEY mode would be dominated by
spectra of grain boundaries. On the other hand, the TFY mode has a much
deeper probing depth ($\sim$100 nm) than the thickness of grain
boundaries, signals from the interior of grains would
be dominant in the spectra.\\
\begin{figure}
  \begin{center}
   \includegraphics[width=86mm]{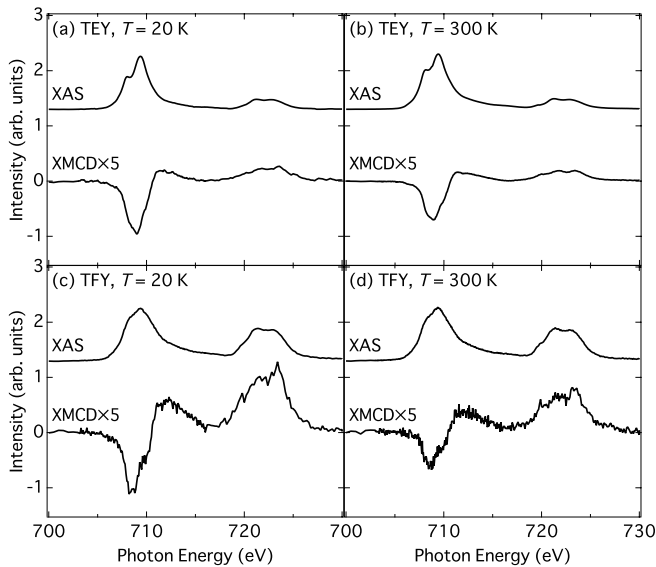}
   \caption{\label{xas}Fe $L_{\rm 2,3}$ XAS and XMCD spectra of
   Sr$_{2}$FeMoO$_{6}$ polycrystal. (a),(b): Spectra taken in the TEY mode at 20
   K and 300 K, respectively. (c),(d): The same as (a) and (b) but taken in the TFY mode. All the spectra
   have been normalized to the Fe $L_{\rm 3}$ XAS peak intensity.}
  \end{center}
 \end{figure}
\indent In Fig. \ref{xas}, we show the XAS [$(\mu_{+}+\mu_{-})/2$] and XMCD ($\Delta\mu = \mu_{+} - \mu_{-}$)
   spectra at the Fe $L_{\rm 2,3}$ (2$p_{\rm 1/2,3/2} \rightarrow$
   3{\it d}) absorption edges normalized to the $L_{{\rm 3}}$ XAS peak intensity. Here, $\mu_{+}$ and $\mu_{-}$ denote the absorption coefficients
   for the photon helicity parallel and antiparallel to the Fe 3$d$
   majority spin, respectively. In these spectra, the higher and lower
   energy parts of the Fe $L_{\rm 2,3}$ peaks are expected to originate
   mainly from the
   Fe$^{3+}$ and Fe$^{2+}$ components in the ground state,
   respectively. The Fe$^{2+}$ signals may also originate from the
   screening of the core hole for the Fe$^{3+}$ ground state. If we compare the spectra
   taken at 20 and 300 K, the XMCD intensity decreases with increasing
   temperature by about 30 {\%} as can be seen from Figs. \ref{xas}(a) and \ref{xas}(b),
   which reflects the decrease of the ferromagnetic moment with
   increasing temperature. The XMCD spectra taken in the bulk-sensitive TFY mode [Figs. \ref{xas}(c) and \ref{xas}(d)] show the same trend but the
   intensity of the Fe $L_{\rm 3}$ peak relative to the $L_{\rm 2}$ peak is
   lowered due to self-absorption and quantitative analysis is
   hindered. Nevertheless, the TFY spectra clearly show a higher Fe$^{2+}$-like
   intensity than the TEY ones, indicating that the grain interior has more
   Fe$^{2+}$ component or stronger core-hole screening due to the higher density
   of conduction electrons than the grain boundaries. In addition, the
   XMCD intensity of the Fe $L_{\rm 2}$ edge in the TFY mode, which is
   less affected by the self-absorption, is higher than that in the TEY
   mode, indicating that the grain interior has
   a larger magnetic moment than the grain boundaries. This suggests that Fe atoms inside the grains are almost fully spin polarized with insignificant amount of non-magnetic component. The XMCD
   intensity in the TFY mode also decreases with increasing
   temperature, corresponding to the decrease of magnetization in the
   grain interior \cite{Kobayashi1998Room-temperatur}.\\
\begin{figure}
    \includegraphics[width=86mm]{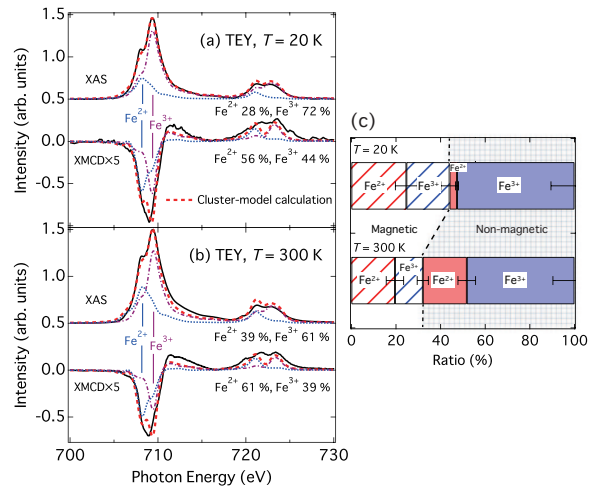}
   \caption{\label{cluster}(Color online) Line-shape analysis of the Fe
   $L_{\rm 2,3}$ XAS and XMCD spectra of Sr$_{2}$FeMoO$_{6}$ polycrystal
   using
   cluster-model calculation. (a) 20 K (b) 300 K. Each of the
   calculated Fe$^{2+}$
   and Fe$^{3+}$ spectra is shown. Vertical bars indicate their $L_{\rm
   3}$ peak positions. The Fe valence ratios deduced from each spectrum
   are also indicated. (c) Ratios of the magnetic and
   non-magnetic components at 20 and 300 K within the probing depth
   ($\sim$5 nm) of the TEY detection. Each component is decomposed into
   the Fe$^{2+}$-like and Fe$^{3+}$-like components. Error bars in each
  component are also indicated.}
\end{figure}
\begin{figure}
 \begin{center}
 \includegraphics[width=86mm]{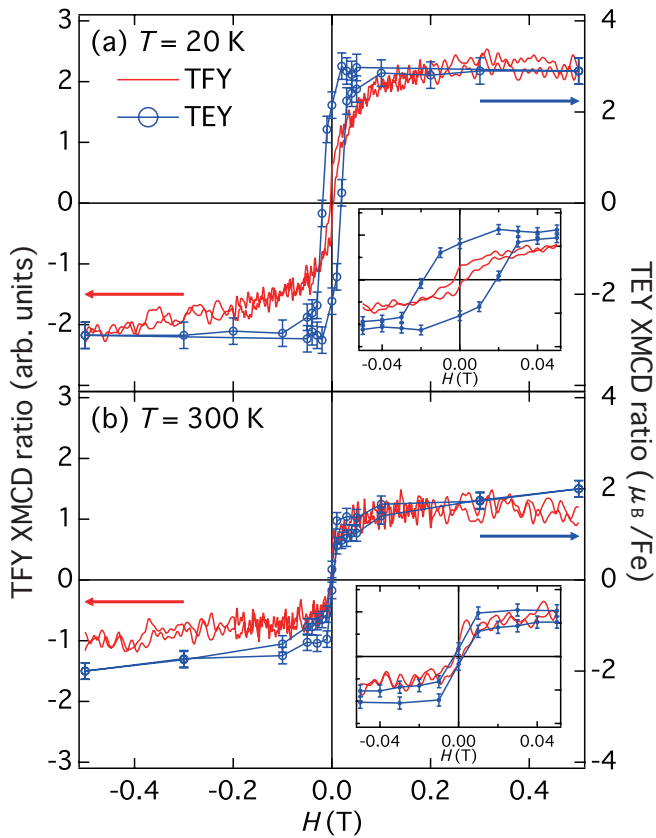}
 \caption{\label{hysteresis}(Color online) Hysteresis loops of the
 Fe $L_{\rm 3}$ XMCD intensities of Sr$_{2}$FeMoO$_{6}$ polycrystal
  taken in the TEY and TFY modes at 20 K (a) and 300 K (b). Inset shows
  an expanded view near $H=0$. The vertical scales of the inset are arbitrary.}
\end{center}
\end{figure}
\indent From now on, we focus on the grain boundary region within the TEY
   probing depth, where the ferromagnetism is weaker than the interior
   of the grains, in order to obtain the quantitative information about
   the Fe valence and magnetization. We performed line-shape
   analysis of the experimental XAS and XMCD spectra by least mean
   square fitting to a superposition of the Fe$^{2+}$ and Fe$^{3+}$
   spectra calculated using the cluster model, as shown by the dotted curves in Fig. \ref{cluster}. Parameters used in this calculation are as
 follows (in units of eV): Fe$^{2+}$: $\Delta=6.0$, $V_{\rm
 E_{g}}=2.1$, $10Dq=0.5$, $T_{\rm pp}=1.2$; Fe$^{3+}$: $\Delta=3.0$, $V_{\rm
 E_{g}}=2.5$, $10Dq=0.5$, $T_{\rm pp}=0.7$. Figures \ref{cluster}(a) and
   \ref{cluster}(b) show that the Fe valence deduced from the XAS spectrum
   is quite different from that deduced from the XMCD spectrum. This
   indicates that the apparent valence of the ferromagnetic Fe ions is more
   Fe$^{2+}$-like while that of the non-magnetic Fe ones is more
   Fe$^{3+}$-like.\\
\indent The experimental XMCD intensities are lower than those expected for fully spin-polarized Fe$^{2+}$ and
Fe$^{3+}$ components, indicating that part of the Fe$^{2+}$ and
   Fe$^{3+}$ ions are in the paramagnetic or antiferromagnetic state. We
   have obtained the weight of the Fe$^{2+}$-like and Fe$^{3+}$-like components in each of the
ferromagnetic and non-magnetic (i.e., para/antiferromagnetic) components, as shown in
Fig. \ref{cluster}(c). In going from 20 to 300 K, the amount of the
ferromagnetic component decreases by 12 \%, in agreement with the
				temperature dependence of magnetization
   \cite{Kobayashi1998Room-temperatur}. Therefore, the ferromagnetic component of the grain boundary is not much different from the ferromagnetic bulk material. The amount of the Fe$^{2+}$-like component
				 in the non-magnetic component
				 increases with increasing temperature,
				 which may partly be due to the actual increase of Fe$^{2+}$ but also due to the
	enhancement of screening effect related with the decrease of
	the resistivity with increasing temperature
	\cite{Kobayashi1998Room-temperatur}.\\
\indent Figure \ref{hysteresis} shows $M$-$H$ curves obtained from the
 XMCD intensity in both detection modes. Hereafter, $H_{{\rm c}}$ of the hysteresis loop of the
 TEY mode and that of the TFY mode are denoted by $H_{{\rm
 c,TEY}}$ and $H_{{\rm c,TFY}}$, respectively. Figure
 \ref{hysteresis}(a) shows that at 20 K $H_{{\rm
 c,TEY}}$ is 0.02 T, 6 times larger than $H_{{\rm
 c,TFY}}$ of 0.003 T. The difference between $H_{{\rm
 c,TEY}}$ and $H_{{\rm c,TFY}}$ is similar to the different $H_{\rm
 c}$ values between the magnetization and MR measurements
 \cite{Sarma2007Intergranular-M}. Because TEY and TFY have the probing depths
 of several nm and $\sim$100 nm, respectively, $H_{{\rm c,TEY}}$ and $H_{{\rm c,TFY}}$ correspond to the $H_{\rm
 c}$ of grain boundaries and that of grains, respectively. This is consistent
 with the scenario that SVMR is
 driven by the large coercivity of the grain boundaries at low
 temperatures \cite{Sarma2007Intergranular-M}. At 300 K, $H_{{\rm
 c,TEY}}$ is reduced and becomes as small as $H_{{\rm c,TFY}}$ as shown in
 Fig. \ref{hysteresis}(b), resulting in the absence of coercivity difference between TEY and TFY, that is,
 between the grain boundaries and the grain interior. This explains the disappearance of SVMR at
 high temperatures \cite{Ray2011Origin-of-the-u}.\\
\indent In conclusion, we have performed a detailed XAS and XMCD studies
of polycrystalline SFMO. The different magnetic hardness of the grain
boundaries and the grains were observed at 20 K as different $H_{\rm
c}$ values in the hysteresis loops recorded in the surface-sensitive TEY and
bulk-sensitive TFY modes. These results are consistent with the SVMR
mechanism proposed for the unusual MR in SFMO polycrystals. From the
line-shape analysis of the XAS and XMCD spectra, we revealed that in the
grain boundary region the Fe valence in the magnetic component is more
Fe$^{2+}$-like than that in the non-magnetic component.\\
 \indent The authors thank S. Ray for useful discussions. This work was
 supported by a Grant-in-Aid for Scientific Research, Basic Research S
 (S22224005) from JSPS. YT was supported by JSPS through the Program for Leading Graduate Schools (MERIT). BP and DDs were supported by the Department of
Science and Technology, Government of India.
%
\end{document}